**Low-temperature electroluminescence excitation mapping of excitons and trions in short-channel monochiral carbon nanotube devices**

Marco Gaulke[1,2], Alexander Janissek[1,2], Naga Anirudh Peyyety[1,2] Imtiaz Alamgir[1], Adnan Riaz[1,2], Simone Dehm[1], Han Li[1], Uli Lemmer[3,4], Benjamin S. Flavel[1], Manfred M. Kappes[1,5], Frank Hennrich[1], Li Wei[6], Yuan Chen[6], Felix Pyatkov[1,2], Ralph Krupke[1,2]

[1] Institute of Nanotechnology, Karlsruhe Institute of Technology, Germany

[2] Institute of Materials Science, Technische Universität Darmstadt, Germany

[3] Light Technology Institute, Karlsruhe Institute of Technology, Germany

[4] Institute of Microstructure Technology, Karlsruhe Institute of Technology

[5] Institute of Physical Chemistry, Karlsruhe Institute of Technology, Germany

[6] School of Chemical and Biomolecular Engineering, The University of Sydney, Australia

Single-walled carbon nanotubes as emerging quantum-light sources may fill a technological gap in silicon photonics due to their potential use as near-infrared, electrically-driven, classical or non-classical emitters. Unlike in photoluminescence, where nanotubes are excited with light, electrical excitation of single-tubes is challenging and heavily influenced by device fabrication, architecture and biasing conditions. Here we present electroluminescence spectroscopy data of ultra-short channel devices made from (9,8) carbon nanotubes emitting in the telecom band. Emissions are stable under current biasing and no quenching is observed down to 10 nm gap size. Low-temperature electroluminescence spectroscopy data also reported exhibits cold emission and linewidths down to 2 meV at 4 K. Electroluminescence excitation maps give evidence that carrier recombination is the mechanism for light generation in short channels. Excitonic and trionic emissions can be switched on and off by gate voltage and corresponding emission efficiency maps were compiled. Insights are gained into the influence of acoustic phonons on the linewidth, absence of intensity saturation and exciton-exciton annihilation, environmental effects like dielectric screening and strain on the emission wavelength, and conditions to suppress hysteresis and establish optimum operation conditions.

## Introduction

The most valuable asset of single-walled carbon nanotubes for photonics applications are their structure-dependent optical transitions, which can be optically or electrically stimulated to emit light in the near-infrared wavelength range.[1,2] This, together with the compatibility of nanotubes to a range of biological, chemical and CMOS processing methods, makes nanotubes highly attractive for applications as novel fluorescence markers in photoluminescence microscopy or as nanoscale emitters for on-chip data transmission with light.[3,4] For devices, the ability to scalably integrate specific nanotubes into complex architectures is essential, and the progress in synthesis and sorting of nanotubes and in selective placement has eventually materialized into electrically-driven, ultra-fast on-chip light emitting devices, that are susceptible to photonic engineering and at the verge of single-photon emission.[5,6] Short channel lengths are required for keeping the device footprint small and operation frequency high. However, if the channel length becomes comparable to the exciton diffusion length, the question arises at which point the emission will be quenched by nearby metal electrodes.[7] So far electroluminescence from single tubes was explored for channel lengths down to a few hundreds of nanometers,[8,9,10,11] or microns for aligned multi-tube devices and thin film devices.[12,13,14] In this work we study emission from channels with an order of magnitude smaller gap size. For such small gaps the question arises, whether emission will be broadened due to large electric fields and generation of hot carriers. This was reported in the pioneering works of the IBM group for micrometer size gaps,[15,8] and is still unexplored for devices with channel length below 100 nm. Furthermore, whether electroluminescence in short channel devices would be driven by impact excitation[16] or carrier recombination[17] and how steady operating points for efficient light generation can be reached is an open question as well. In this work we have fabricated devices with a channel length down to ten nanometers and recorded electroluminescence spectra under controlled biasing conditions over a wide temperature range. We studied preferentially monochiral (9,8) nanotubes because they emit in the technologically relevant telecom band and can be synthesized through selective catalyst CVD growth.[18,19] At cryogenic conditions we observe excitonic and trionic emissions that are exceptionally stable and reproducible, appearing at well-defined source-drain current and gate voltage. We have compiled electroluminescence excitation maps which allow identifying optimum operating conditions, where emission lines are narrow and device efficiency is high. The enhanced stability is the result of a specific sample process flow, device biasing scheme, vacuum and low temperature. We emphasize that the data constitutes the first electroluminescence spectra recorded at cryogenic conditions. The results are important for understanding the mechanism behind light emission from short channel devices and to advance their potential use as on-chip light sources.

**Results and Discussion**

(9,8) nanotube devices with Pd source-drain electrodes were fabricated on 300 nm $SiO_2$ / p-doped Si substrates by electron-beam lithography, metallization and electric-field assisted deposition of nanotube (dielectrophoresis). The nanotubes were produced by selective-catalyst CVD, dispersed in toluene by polymer-wrapping, and purified and length sorted by gel filtration (details in methods section). Absorption spectra (Fig. 1a) and photoluminescence excitation maps (Fig. 1b) of excitonic transitions within the telecom band give evidence for a high content of (9,8) nanotubes and the presence of (8,7), (9,7), (10,8), (10,9) species in minor concentrations. To comply with the nominal charge transfer length for side-contacted nanotubes,[20] fractions of length-sorted (9,8) nanotubes were selected for depositions such that nanotubes were at least 200 nm longer than the distance between the source-drain electrodes. Simultaneous site-selective deposition of single-tubes from diluted nanotube dispersions onto multiple electrode pairs (Fig. 1c), with channel lengths between 10 - 1000 nm, was carried out by dielectrophoresis as described in methods section. A representative contact is shown in Figure 1d+e. The devices were electrically wired, mounted into an optical cryostat and evacuated to $10^{-7}$ mbar. The cryostat is part of an optical microscopy and spectroscopy setup to image and analyze light emanating from devices with diffraction-limited spatial resolution and 2 nm spectral resolution (1.2 meV @ 1440 nm). Electroluminescence was measured in the application relevant telecom band from 1200 to 1610 nm and in the temperature range $4 - 300$ K. All spectra were corrected by the relative spectral sensitivity of the setup. In the following we will report on electroluminescence spectra recorded from ultra-short channel devices at cryogenic temperatures, which to the best of our knowledge are the first of its kind.

Figure 2a and 2b show $E_{11}$ emission at room temperature from (9,8) devices with 50 nm and 100 nm gaps, and the corresponding linewidths for a number of devices with gap sizes between 10 to 1000 nm are shown in Figure 2c. The fact that we could measure electroluminescence down to 10 nm gap size was surprising, because from photoluminescence studies it is known, that the exciton diffusion length in nanotubes at room temperature is on the order of hundreds of nanometers.[21] It is also known that photoluminescence is quenched for fluorophores in direct contact with metallic surfaces.[7] Therefore, if the device channel length is short compared to the nanotube exciton diffusion length, one would expect that the non-radiative recombination of excitons at the nearby source and drain metal contacts suppresses electroluminescence. In particular for ultra-short channel devices with gaps between 10 to 50 nm we would have expected that electroluminescence becomes difficult to observe. On the other hand, the Strauf group recently reported plasmonic-enhanced photoluminescence from nanotubes located on top of metallic bow-tie antennas, having similar gap dimensions.[22] A dielectric spacer was added on top of the metal structures to avoid quenching of photoluminescence. Such a layer is absent in our experiment. We might speculate that in our case the polymer shell around the nanotubes acts as sufficiently thick dielectric spacer. Based on simulations of Wang et al.,[23] we can assume that the polymer shell has a thickness of ~1 nm, in agreement with the absence of a massive electrical tunneling barrier which appears for thicker polymer shells. For fluorophores, 1 nm distance to a metal surface would be far too small to avoid quenching[7] . On the other hand, for nanotubes on metal surfaces, Hong et al. have shown that the quenching distance d is much smaller and well below 3 nm.[24] How can this be understood? Barnes has pointed out that the distance dependence of the decay rate follows $d^{-3}$ for non-radiative decay into bulk modes.[25] This occurs if at the emission frequency electron scattering in the metal is strong, and it has been shown to be the case for fluorophores emitting in the visible.[26] However in the near-infrared, where electron scattering is reduced and decay into bulk modes is suppressed, the decay into non-radiative surface plasmon polariton modes (SPP) prevails and a $d^{-4}$ dependence is expected.[27] The exponent can be understood by the dipole-dipole Förster energy transfer rate, which depends inversely on the sixth power of the distance:[26] For the case of a dipole above a metal, the problem is equivalent to one in which a point dipole transfers energy to a volume of point dipoles. The rate must be integrated over this volume and the distance dependence is thereby reduced to cubic. From the same consideration one expects quartic distance dependence for transfer to a surface or thin film. Hence for near-infrared active (9,8) nanotubes it is understandable that the quenching distance is much smaller as for fluorophores emitting in the visible. On the other hand, electroluminescent light decaying into SPPs might also be recovered by the nanometer gap sized electrodes converting SPPs back to photons that couple to the far field.

Of course one could also envision a suppression of quenching by nanotube defects present in the gap region, in analogy to the defect-induced reappearance of photoluminescence in ultra-short nanotubes.[28] This is however unlikely to be the case here, since the electroluminescence data does not show the characteristic red-shifted emission of an $sp^3$-functionalized nanotube.[28] From a technological point of view it is very promising that electroluminescence from carbon nanotubes is not bound to devices with large channel length. Also, the linewidth does not significantly broaden towards smaller gap size (Fig. 2c), despite the very large induced electric field within the gap of up to $10^8$ V/m.[29] Compared to the early electroluminescence work of the IBM group,[8] the linewidth is more than an order of magnitude narrower and thermalization of hot carriers appears to be sufficient to sustain cold emission even in ultra-short channel devices.

We discuss now the temperature dependence of the excitonic emission of (9,8) devices. Figure 3a shows a set of electroluminescence spectra of a representative device measured between 4 and 300 K. The $E_{11}$ emission, visible at ≈1440 nm, continuously narrows down from 50 nm (30 meV) linewidth at 300 K to 5 nm (3 meV) linewidth at 4 K. This dependency also holds for other (9,8) devices at moderate current biasing (< 100 nA), as shown in Figure 3b. The lineshape is Lorentzian throughout the entire temperature range. Since the linewidth is larger than the spectral resolution of our setup, we conclude that the homogeneous broadening is caused by exciton dephasing.[30] Following the temperature-dependent (9,8) photoluminescence data of Yoshikawa et al.,[31] we assume that the temperature dependent part of the linewidth broadening is due to coupling to acoustic phonons. We fitted the linewidth $\Gamma$ in Figure 3b to the expression $\Gamma = \Gamma_0 + A \cdot T + B \cdot (\exp(E_{ph}/k_B T)-1)^{-1}$ .[32] Coefficients A and B are the exciton-phonon coupling constants for acoustic and optical phonons, respectively, and $E_{ph}$ is the optical phonon energy. The temperature dependent electroluminescence linewidth data scatters around a straight line with A ≈ 0.085 meV/K and B = 0, which indicates that the coupling to low-frequency acoustic phonons determines the increase of the linewidth with temperature, and that contributions of the high-frequency modes are negligibly small, as observed by Yoshikawa et al. in photoluminescence.[31] Interestingly we see that the Lorentzian linewidth increases with the current bias and reaches up to 100 meV at 300 nA and at 4 K (Fig. S1a-b). This is comparable to previous high-bias electroluminescence measurements of (9,7) tubes, measured at room temperature, where exciton-exciton annihilation (EAA) was hold responsible for a reduction of the lifetime with increasing current.[33] However, we observe here that the emission intensity increases linearly with the current (Fig. S1d) and not sublinear as expected for an EAA limited emission rate.[34] Hence we conclude that EAA is not important here and consider dephasing by exciton-exciton scattering as a source for current induced line broadening.[35]

Concerning the residual linewidth $\Gamma_0$ at 4 K, we have measured on average between 3 to 5 meV with champion devices down to 2 meV (Fig. 3b+c). The (9,8) linewidth at 4 K obtained by electroluminescence is therefore comparable to (9,8) photoluminescence data.[31,36] In some cases, albeit for other chiralities, photoluminescence linewidth was reported to reach 1 nm or below, with significant variations from tube to tube.[37,38] These works indicate a sensitivity of the linewidth to the confinement of the low-energy acoustic phonon modes and to exciton localization.[39,38] Variations in the environment can easily induce these effects and device-to-device variation of $\Gamma$ is likely a result of limited precision in the directed placement of nanotubes, variations in the ordering of the polymer around the nanotube and insufficient interface engineering on the atomic scale. Likewise we observe variations in the temperature dependence of the $E_{11}$ peak position from device to device and with current (Fig. 3d). Concerning the peak position, we observe for low currents a redshift up to 17 nm (10 meV) with increasing temperature from 4 K to 300 K. Sign and magnitude of the shift match with the expected temperature dependence of the nanotube bandgap.[40] We also observe a blueshift at intermediate temperatures, which we may interpret as strain-induced as reported for polymer-wrapped $\nu$ = (n-m)mod3 = 1 nanotubes.[41] The accumulation and the release of interfacial strain probably vary from device to device and may explain why blueshifts do not occur at identical temperatures. Also, we cannot exclude the effects of water that might have been trapped inside the nanotubes during device fabrication and as a result phase transitions in the orientation of water dipoles may contribute to the observed wavelength shifts as well. Molecular dynamics simulations show that the alignment of the dipoles and the freezing temperature is diameter dependent[42,43] and occurs for suspended (9,8) nanotubes between 250-290 K. Since the high-frequency permittivity of frozen water is smaller than liquid water, the dielectric screening changes accordingly.[43,44] Hence the redshift in the upper temperature range might also be caused by melting of ice inside the carbon nanotube.

We report now on the influence of the back-gate voltage on the emission characteristics in short channel devices. Figures 4a shows that for moderate current levels (< 100 nA) the excitonic emission can be completely suppressed when switching from -2 V gate voltage to +4 V. At the same time a new peak emerges at ≈1585 nm, which for a fixed current bias can be gradually switched on to a peak intensity comparable to the excitonic peak (Fig. 4b). This behavior is typical, and shown here for two (9,8) devices. We have measured similar dependencies also for other chiralities and determined the energy difference $\Delta E$ between the excitonic emissions and the corresponding red-shifted emissions. Figure 4c shows that the energy difference decreases with the nanotube diameter $d$ and fits to $\Delta E = A/d + B/d^2$ with A = 40±10 meV/nm and B = 65±9 meV/nm$^2$. The data reproduces nicely the energy difference between excitons and trions observed by Park et al. in the photoluminescence of electrochemically doped nanotube films,[45] and by Jakubka et al. in the electroluminescence of thin film devices.[14] We are therefore confident, that the red-shifted emission in the electroluminescence spectra of short-channel devices under positive gate voltage stems from the recombination of trions. Trions in nanotubes were first reported by Matsunaga et al. in the photoluminescence of hole doped nanotubes in solution,[46] explaining that the energy difference observed between the excitonic and trionic emission equals the sum of the trion binding energy ($\propto d^{-1}$) and the single-triplet exciton exchange splitting ($\propto d^{-2}$); in agreement with theory.[47] Figure 4c shows a comparison of our data with the other works. We note that the binding energy and the exchange energy, and hence A and B, depend on the dielectric environment of the nanotube,[44] and slight variations between experiments are expected. Plotting the trion emission energy against the exciton emission energy (Fig. 4d), we find a linear correlation as expected from the diameter dependence of the excitonic emission,[44] and the diameter dependence of the energy difference between the excitonic and trionic emission.

For applications it is important to have stable and reproducible operating points for excitonic or trionic emission and it is necessary to describe the conditions and biasing schemes used in this work.

A major source for drift and irreproducibility in short-channel electroluminescence measurements is related to hysteresis in the transconductance curves. Figure S3 shows a typical measurement taken at room temperature (296 K) and under vacuum ($10^{-7}$ mbar). A hysteresis of 7-8 V is prominent and typical for as-prepared short-channel devices. In such cases it is impossible to obtain stable steady-state current and emission at a fixed gate voltage, because the filling and depleting of trap states with time results in an effective time-dependent gating of the nanotube. Upon moderate heating the device to 343 K for 120 min within the evacuated cryostat, we observe a significant reduction of the hysteresis to ~2 V, which reduces to below 1 V at 220 K and disappears below 100 K (Fig. 5a). A more rigid evaluation of hysteresis is to integrate the area between the forward and backward sweeps of the transconductance curves, which shows that a full suppression of hysteresis occurs only at 4 K (Fig. 5b). Water-related trap states are probably the major source of hysteresis since moderate vacuum annealing already significantly reduces hysteresis, and as discussed before, water, that may be encapsulated during device processing, could play a role here as well. For stable light emission it is however also important to apply current biasing instead of voltage biasing. This is because the device resistance $R$ decreases with temperature $T$ ($dR/dT < 0$), as seen from the temperature dependence of the ON-state current (Fig. 5a). Subsequently voltage biasing favors fluctuations in the electrical power dissipation and leads to unstable emission and thermal runaway. In contrast, current biasing stabilizes temperature, reduces power fluctuations and leads to enhanced emission stability. Furthermore – and maybe most important - voltage biased nanotube devices often show unipolarity in the transconductance curves at low source-drain bias, with ambipolarity appearing only at larger bias, as shown in Figure 5c. This behaviour is common for nanotube/metal contacts where the Schottky barrier for electrons is larger than for holes. It impedes finding optimum operating points for light emission since the regions of electron and hole conduction depend on both, gate voltage and source-drain voltage. In contrast, by imposing source-drain current bias, the voltage required for compensating the different resistances in the respective p- and n-regions will be instantaneously applied by the source-meter electronics. This leads to well-defined gate-voltage controlled p- and n-regions and pronounced ambipolarity.

Figure 6a shows the result of such current biasing for a (9,8) device, where the measured voltage across the source-drain electrodes is plotted against the applied source-drain current and gate voltage. The p- and n-regions in the map are marked and separated by a region in which the voltage goes through a maximum at around -10 V gate. This is the region with identical electron and hole currents, which is offset from the origin due to weak n-doping. The asymmetry of the map is as discussed before due to non-identical Schottky barriers for electrons and holes. During the measurement of the map, we have recorded simultaneously electroluminescence spectra. Spectra have been taken with an integration time of 10 seconds for each of the 42 steps in gate voltage and 5 steps in source-drain current, summing up to 35 minutes for the parameter space. The spectra were then integrated in wavelength sections corresponding to the excitonic emission (1370-1500 nm) and the trionic emission (1500-1613 nm). The resulting excitonic and trionic excitation maps are shown in Fig. 6c-d. Different regions can be identified, in which predominantly excitons and trions are formed. Excitons are formed in the region with identical electron and hole currents (Fig. 6c), which is evidence for light generation through carrier recombination and not impact excitation. This efficient mechanism of light generation has been observed previously in long-channel and thin film devices,[17,13] and this work now shows that carrier recombination is also dominating in short-channel devices. The circumstances for trionic emission are somewhat more complicated, since the formation of trions requires a net charging of the nanotube channel, a condition which cannot be satisfied at charge neutrality. If we compare Fig. 6d with Fig. 6c and Fig. 6a, we notice that the gate voltage range for trionic emission is shifted by +10 V against the gate voltage for excitonic emission, and hence is occurring in the region with excess electrons. We note that for n-doped and p-doped devices we observe the corresponding negatively charged trion (T-) and the positively charged trion (T+), respectively. Figure S2 shows an example of a p-doped device with the T+ emission occurring in the p-region (with excess of holes) at more negative gate voltages with respect to the excitonic emission. To identify conditions of enhanced emission efficiency, we have normalized the electroluminescence excitation maps Fig. 6c and Fig. 6d with the electrical power dissipation map Fig. 6b. The obtained Figures 6e-f yield maps of relative power efficiency for excitonic and trionic emission given as count rate per electrical power. We observe that the efficiency for excitonic emission peaks around -10 V gate and is rather independent from the current, whereas the trionic emission at +10 V gate becomes efficient only at larger current bias. We are lacking an explanation for this observation but as shown in Fig. S2 and Fig. 4a-b there are also examples where trion emission occurs already at very low current bias.

We also determined the electroluminescence quantum efficiency $\eta_{ELQE} = N_{photons} / N_{charges}$, the ratio of emitted photons and charges passing through the nanotube, which requires knowing the sensitivity of the setup to photons emitted by the nanotube. We approached the problem by calculating analytically the radiation pattern of an emitter on a layered substrate,[48] the fraction of emitted photons collected by the microscope objective, and determined experimentally the factor that converts detector count rate into photon flux. The product of both yields the total setup efficiency $\eta_{setup}$ and converts detector count rate to photon flux for the specific experimental setup with a nanotube on a 300 nm-SiO$_2$/Si substrate. The spectra shown here were already corrected for the relative spectral sensitivity of the setup and for $\lambda > 1300$ nm the conversion into photon flux is then achieved by division with the factor $\eta_{setup} = 8.7 \cdot 10^{-5}$ counts·s$^{-1}$/ photons·s$^{-1}$. Details and procedures are described in the supporting information. We can directly calculate $\eta_{ELQE}$ for excitonic and trionic emission via

$$\eta_{ELQE} = \frac{N_{photon}}{N_{charges}} = \frac{\text{intensity(cps)}}{\text{current(A)}} \frac{2e}{\eta_{setup}}$$

With $\eta_{ELQE} = 3 \cdot 10^{-6} \cdot \text{intensity[cps]/current[nA]}$ we obtain for devices at the optimum operating point $\eta_{ELQE} = 5 \cdot 10^{-4}$, which reproduces the result from waveguide-coupled (9,7) nanotube emitters.[5]

## Conclusion

In summary, we have shown that electroluminescence in ultra short-channel devices is not quenched, likely due to relaxed quenching distance constraints in the near-infrared. We have also realized on / off-switching of excitonic and trionic emission by the gate voltage. Such control over the emission is important for the development of reliable and stable on-chip light sources with narrow-line emission in the telecom band. The first cryogenic electroluminescence spectroscopy data shown in this work gives new insights into the mechanism and the limitations of electrically-induced light emission: The temperature dependence of the electroluminescence linewidth is in line with exciton dephasing caused by low-energy acoustic phonon, whereas the homogeneous line broadening with increasing current at constant efficiency indicates dephasing by exciton-exciton scattering. The observed linewidth of ~2 meV at 4 K also shows that cold electroluminescence prevails even in short-channel devices, which is promising for applications. Current biasing enforces ambipolarity and leads to stable, gate-voltage only controlled operating points for light emission. By recording electroluminescence excitation maps we could verify that light emission in short-channel devices is generated by carrier recombination. For applications it will be crucial to achieve the degree of control also at room temperature.

## Methods

*Device fabrication*: Devices were prepared from commercial substrates (Active Business Company), which consist of a boron doped silicon carrier wafer (resistance $\Omega < 0.005$ cm) covered with 300 nm of thermal silicon oxide. The wafer was diced to 10x10 mm² to fit into the cryostat setup and electrodes were defined by electron beam lithography (Leo 1530) with proximity correction. While the preparation of structures with channel length of 150-1000 nm involved standard e-beam patterning, the ultra-short channel devices down to 10 nm electrode gap size were fabricated as followed. Samples were cleaned with acetone, isopropanol and oxygen plasma, and spincoated with 30 nm thick positive resist polymethyl methacrylate (PMMA 950K 1% in Anisol). After e-beam patterning the sample was developed in a solution of MIBK and isopropanol (1:3, for 30 s at 0 °C) and annealed on a hot plate for 60 s at 90 °C. 5 nm of chromium and 25 nm of palladium were deposited by sputtering technique. The lift-off procedure was performed in acetone under mild sonication.

*Preparation of CNT-suspensions and CNT deposition*: SWCNT was synthesized using $CoSO_4/SiO_2$ as a catalyst and CO as a carbon precursor as described in detail.[49] The catalyst (200 mg) loaded in a 1-inch tubular reactor was first reduced under $H_2$ flow (1 bar, 50 sccm) while the reactor temperature was increased to 540 °C. Then, the reactor temperature was further increased to 780 °C under Ar flow (1 bar, 50 sccm). Afterward, the catalyst was exposed to CO (6 bar, 100 sccm) to initiate SWCNT growth for one hour. Raw SWCNT soot was obtained after dissolving $SiO_2$ in the catalyst loaded with SWCNTs in NaOH (1 M) solution. The preparation of CNT suspensions used in this study is described in full detail in our previous works.[50] For SWCNT suspensions 100 mg of the raw SWCNT soot and 100 mg of the polymer poly(9,9-di-ndodecylfluorenyl- 2,7-diyl) (PODOF) (Sigma-Aldrich) were mixed in 100 mL of toluene and subjected to a sonication treatment for 2 h by using a titanium sonotrode (Bandelin, ~20% power). During sonication, the suspension was placed in a water-circulation bath to aid cooling. After sonication, the suspension was then centrifuged for 2 h at 20000 g. To generate the starting suspensions for size exclusion separation the supernatant was concentrated to ~5 mL by evaporating ~95 mL of toluene. Semipreparative, size exclusion chromatography was performed using Toyopearl HW-75 resin (Tosoh) filled into a glass column having 16 mm inner diameter and 20 cm length. After application of 5 mL of SWCNT starting suspension to the gel, the sample was flowing through the gel under gravity resulting in a flow rate of ~2 mL/min with toluene as eluent. Fractions were collected in ~4 mL portions. UV–vis–NIR absorption spectra of the fractions were recorded on a Varian Cary 500 spectrophotometer. Photoluminescence maps were measured in the emission range of ~900–1700 nm and excitation range of 500–950 nm (scanned in 3 nm steps) using a modified FTIR spectrometer (Bruker IFS66) equipped with a liquid-nitrogen-cooled Ge-photodiode and a monochromatized excitation light source as described elsewhere.[51] Toluene-based suspensions contain few-chirality semiconducting nanotubes with diameter of 1-1.2 nm, dominating by (9,8)-CNTs. Individual CNTs were simultaneously deposited from solution onto multiple contact pairs by capacitive-coupled ac-dielectrophoresis.[52] The suspension was diluted by a factor of 10-100 in order to deposit individual tubes and a 20 µl droplet was placed onto the device array. A bias between 0.1 and 2 V at frequencies between 100 kHz and 1 MHz was applied between the common drain electrode and the back gate using Agilent 33250 function generator. After 5 minutes the sample was rinsed with toluene to get rid of the excess polymer and annealed for 1.5 hours at 160 °C in order to improve the contact conductivity. To confirm the deposition of individual CNTs transport characteristics of the devices were measured at ambient conditions in a probe station with TRIAX probes using an Agilent 4155C Semiconductor Parameter Analyzer.

*Electroluminescence spectroscopy and cryogenic setup*: Samples were mounted on a custom made sample holder into a 4-500 K helium-flow, sample-in-vacuum high-resolution microscopy cryostat system (MicrostatHiResII, Oxford). Chip contacts of up to eight devices were bonded onto palladium pads attached to this holder. In-situ annealing at 60-70 °C was conducted via the integrated heater at pressure below $10^{-6}$ mbar and the subsequent electroluminescence measurements were carried out without breaking vacuum. The cryostat has a 10 mm diameter optical access via a 0.5 mm thick quartz window and the emitted light was collected with a Zeiss LD-Plan Neofluar objective (40x/0.6) of a customized Zeiss Axiotech Vario microscope and focused with an off-axis parabolic mirror (Thorlabs MPD149-P01, Ag, 25.4 mm, f/4) into an Acton SP-2360 (f/3.9) imaging spectrograph (Princeton Instruments) and dispersed via a 85 G/mm, 1.35 µm blazed grating onto a InGaAs photodiode linear array (PyLoN-IR Princeton Instruments) with 1024 pixels, sensitive from 950 - 1610 nm. The absolute spectral sensitivity of the setup was calibrated as described in the Supporting Information. The cryostat is positioned with sub-µm precision by a motorized xy scanning stage (8MTF, Standa) and the working distance between objective and sample surface is adjusted by a high precision objective piezo scanner (P-721 PIFOC / E-665 Piezo Amplifier, Physics Instruments), which allowed a precise and stable positioning of the emitter. CNT-devices mounted in the cryostat were driven by Agilent 4155B Semiconductor Parameter Analyzer.

**Associated content**

Supporting Information

Additional data on the high bias dependence of excitonic emission (Fig. S1), electrical biasing and power dissipation versus light emission from excitons and trions in a hole-doped (9,8)-device (Fig. S2), impact of annealing on transconductance curve (Fig. S3), and measurements and simulations regarding the electroluminescence detection efficiency of the setup.

**Author information**

Corresponding Authors

E-mail: krupke@kit.edu

Author Contributions

The experiments were conceived and designed by R.K., M.G., and F.P. The nanotube raw material was provided by L.W., and Y.C., purified by F.H., and H.L. and B.F., and length fractioned by F.H. and M.M.K. Devices were fabricated by M.G., A.J., and S.D. The low-temperature electroluminescence setup was built by M.G. with input from I.A., U.L., F.P. and R.K.. Electroluminescence measurement were performed by M.G. and A.J. with input from F.P. and A.R.. Simulations were performed by N.P. and R.K.. The manuscript was written by R.K., M.G. and F.P. with input from all co-authors.

**Acknowledgements**

A.R., F.P. and R.K. acknowledge funding by the Volkswagen Foundation. N.P., H.L., and B.F. acknowledge funding by the Deutsche Forschungsgemeinschaft (DFG). M.G., F.H., M.M.K., B.F., U.L., F.P. and R.K. acknowledge support by the Helmholtz Society through the program, Science and Technology of Nanosystems (STN), and by the Karlsruhe Nano Micro Facility (KNMF). R.K. acknowledges support by Lukas Novotny for simulating dipole emission, Achim Hartschuh for insights into SPP modes, and Yoshikazu Homma for valuable information about environmental effects on carbon nanotubes.

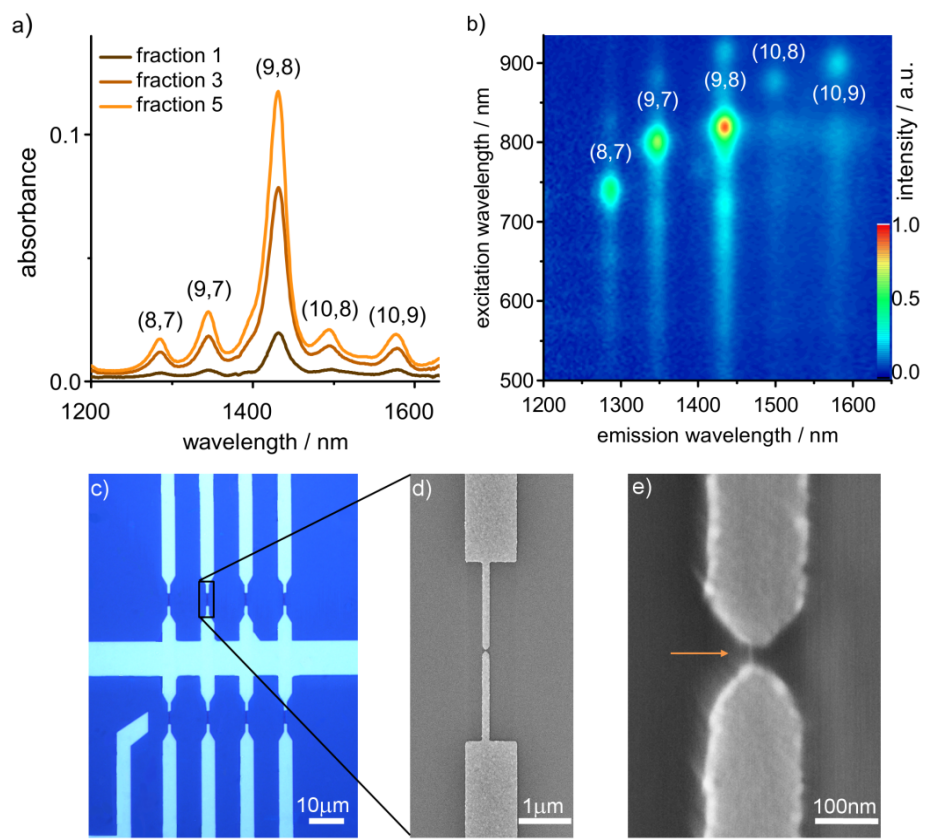

**Figure 1**: (a) Absorption spectra of length-fractionated, polymer-wrapped carbon nanotubes dispersed in toluene containing mainly the (9, 8) chirality. (b) Photoluminescence excitation map of fraction 5. (c-d) Optical image and scanning electron micrograph of the device layout. (e) Typical (9, 8) single-tube contact formed after deposition from solution with dielectrophoresis. The position of the nanotube is indicated.

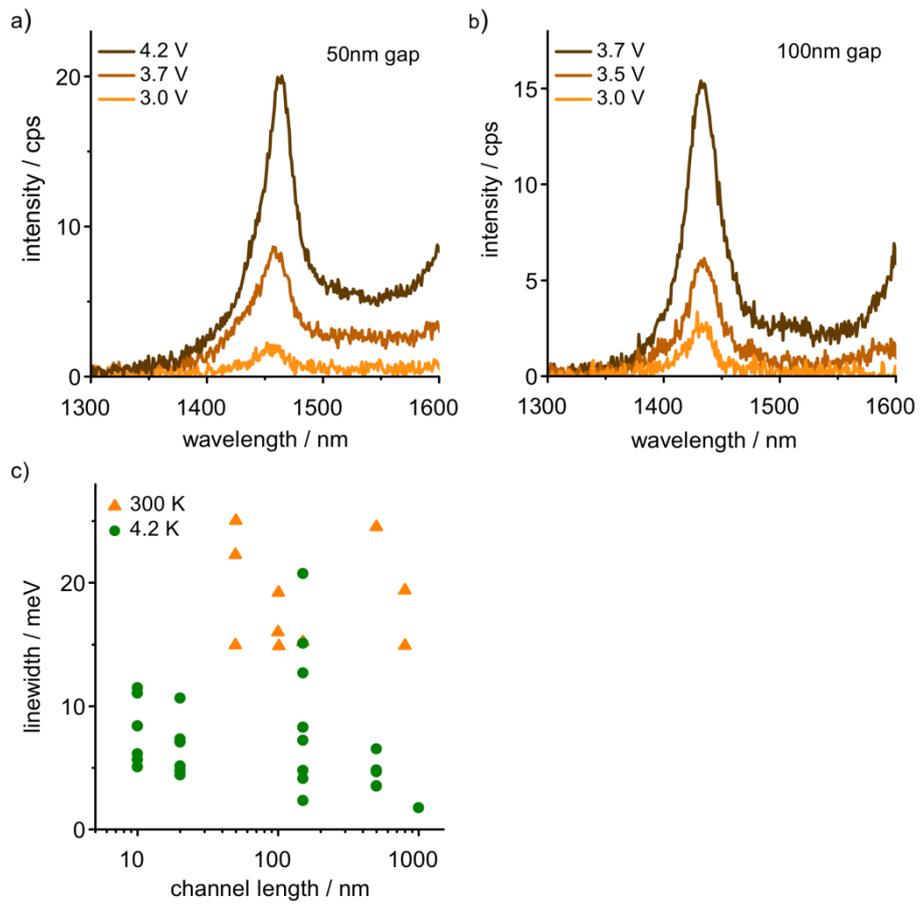

**Figure 2**: Room-temperature electroluminescence spectra of (9,8) short-channel devices with channel length of 50 nm (a), and 100 nm (b), under increasing bias voltage. (c) Full width at half maximum linewidth measured at 300 K and 4 K.

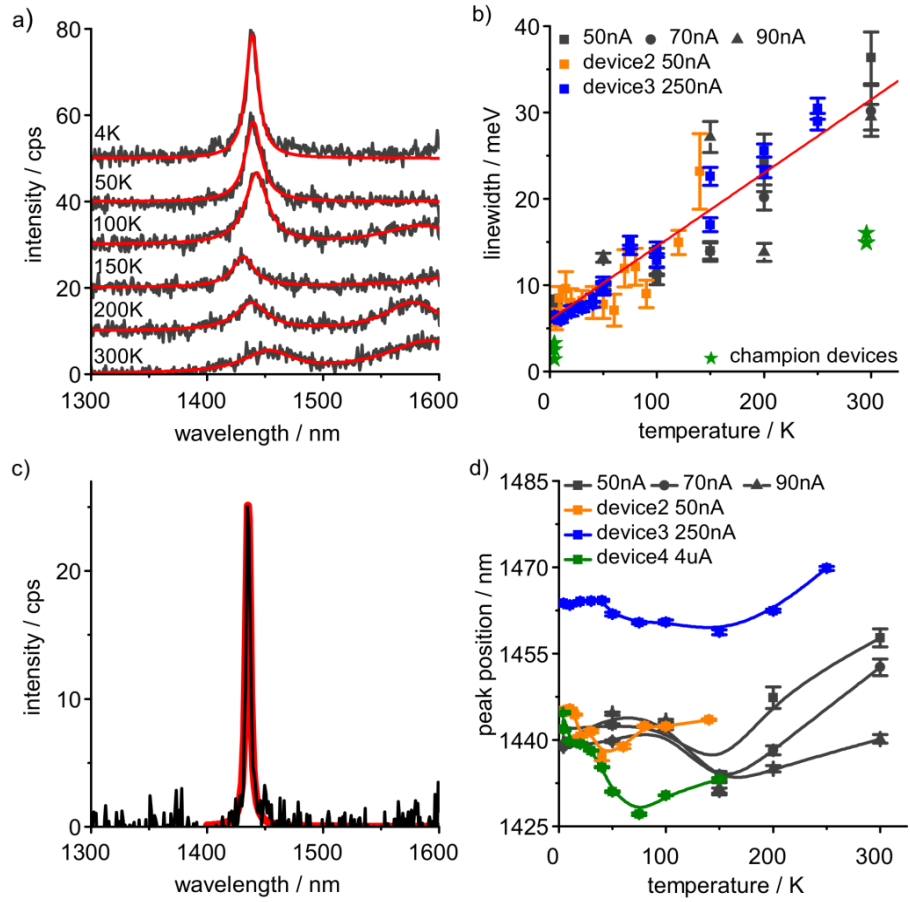

**Figure 3**: Temperature dependence of (9,8) devices: (a) electroluminescence spectra, (b) linewidth, and (d) peak position. (c) Excitonic emission of a "champion device" with 2 meV linewidth at 4 K. Devices were measured at specified current levels. Spectra were fitted with Lorentzian functions with uncertainties in linewidth and peak position given as error bars in (b) and (d).

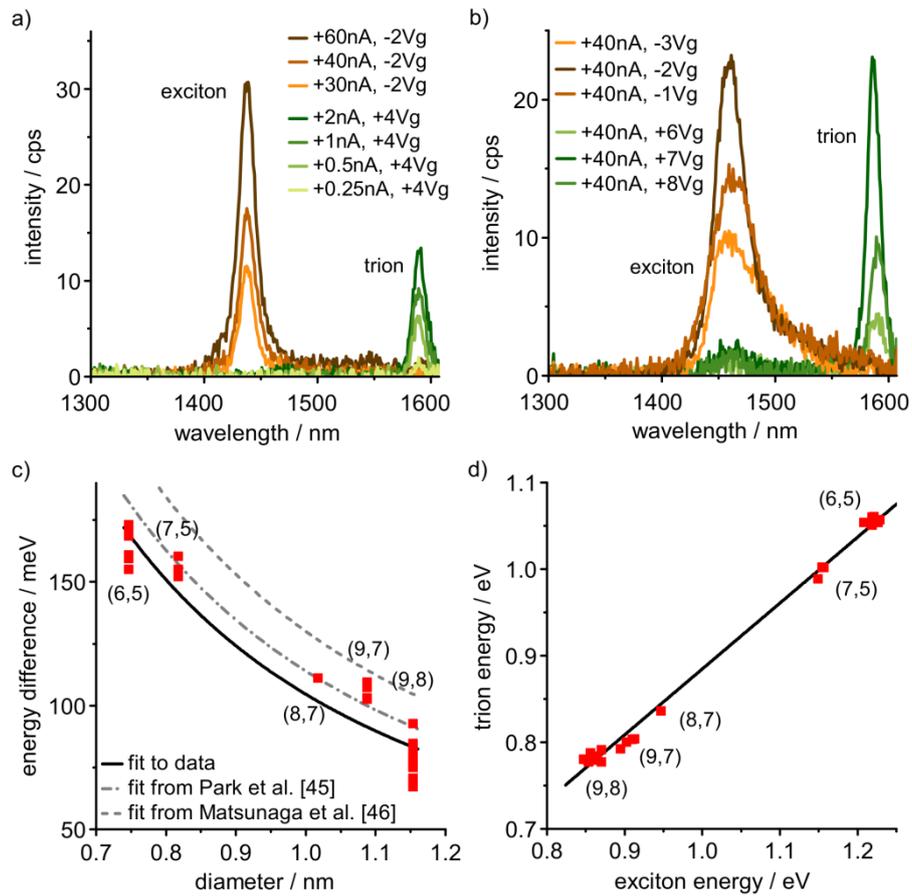

**Figure 4**: Gate-controlled switching between excitonic and trionic emission from (9,8)-devices: (a) spectra acquired at discrete gate voltage and increasing current, and (b) increasing gate voltage and fixed current. (c) Trion emission energy against exciton emission energy measured for several (n,m) devices. (d) energy difference between excitonic and trionic emission versus nanotube diameter. Fits to data discussed in text.

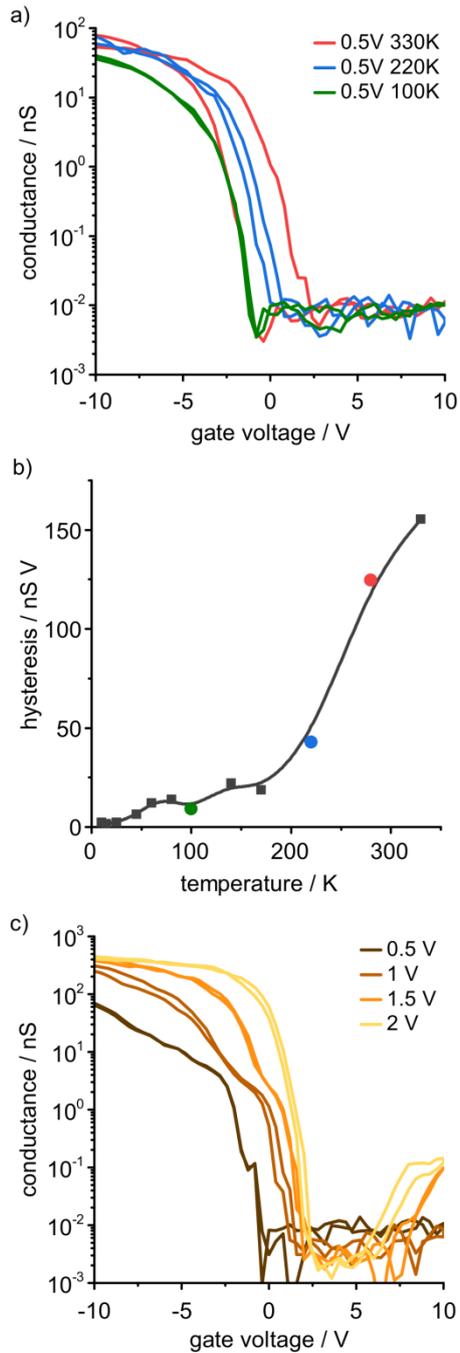

**Figure 5**: Electrical device characteristics: (a) Temperature dependence of the transconductance measured at fixed bias voltage. The hysteresis is defined as the area between the forward and backward sweeping curves and plotted in (b). (c) Transconductance curves at 4.2 K with increasing voltage bias showing negligible hysteresis.

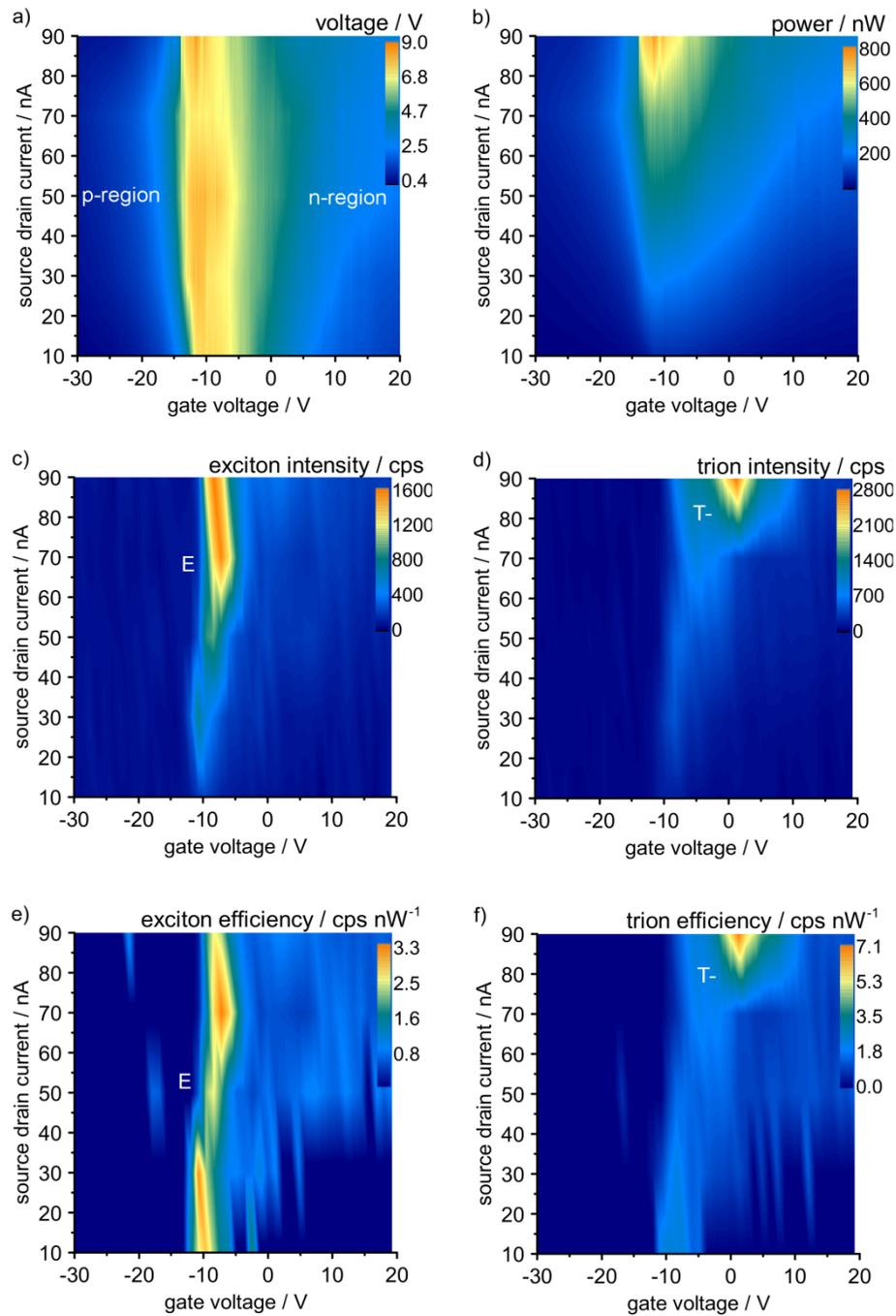

**Figure 6**: Current-driven excitonic and trionic light emission from a weakly n-doped (9,8)-device: (a-d) Source-drain voltage, electrical power, and integrated exciton and trion intensity versus source-drain current and gate voltage (exiton range: 1370-1500 nm; trion range: 1500-1613 nm). Regions of electron (n) and hole (p) conduction are indicated. (e-f) Exciton and trion efficiency in count rate / electrical power. The negatively charged trion (T-) appears in the n-region whereas the excitonic emission (E) appears close to charge neutrality between the p- and n-regions.

**Low-temperature electroluminescence excitation mapping of excitons and trions in short-channel monochiral carbon nanotube devices**

Marco Gaulke[1,2], Alexander Janissek[1,2], Naga Anirudh Peyyety[1,2] Imtiaz Alamgir[1], Adnan Riaz[1,2], Simone Dehm[1], Han Li[1], Uli Lemmer[3,4], Benjamin S. Flavel[1], Manfred M. Kappes[1,5], Frank Hennrich[1], Li Wei[6], Yuan Chen[6], Felix Pyatkov[1,2], Ralph Krupke[1,2]

[1] Institute of Nanotechnology, Karlsruhe Institute of Technology, Germany

[2] Institute of Materials Science, Technische Universität Darmstadt, Germany

[3] Light Technology Institute, Karlsruhe Institute of Technology, Germany

[4] Institute of Microstructure Technology, Karlsruhe Institute of Technology

[5] Institute of Physical Chemistry, Karlsruhe Institute of Technology, Germany

[6] School of Chemical and Biomolecular Engineering, The University of Sydney, Australia

**Supporting Information**

Content: Data on the high bias dependence of excitonic emission (Fig. S1), electrical biasing and power dissipation versus light emission from excitons and trions in a hole-doped (9,8)-device (Fig. S2), impact of annealing on transconductance curve (Fig. S3), and measurements and simulations regarding the electroluminescence detection efficiency of the setup.

**High-bias electroluminescence data**

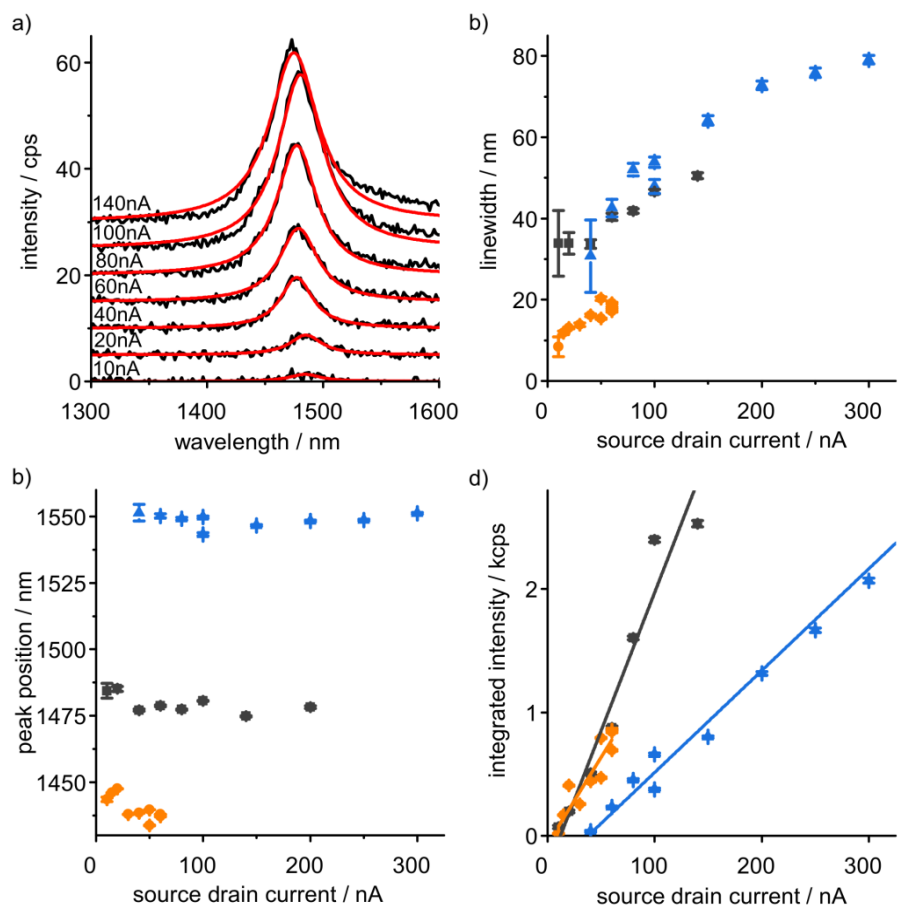

Figure S1: Excitonic emission spectra (a), FWHM-linewidth (b), peak position (c), and integrated intensity (d) of various devices versus source drain current in the high bias regime.

**Light emission from excitons and trions in a hole-doped (9,8)-device**

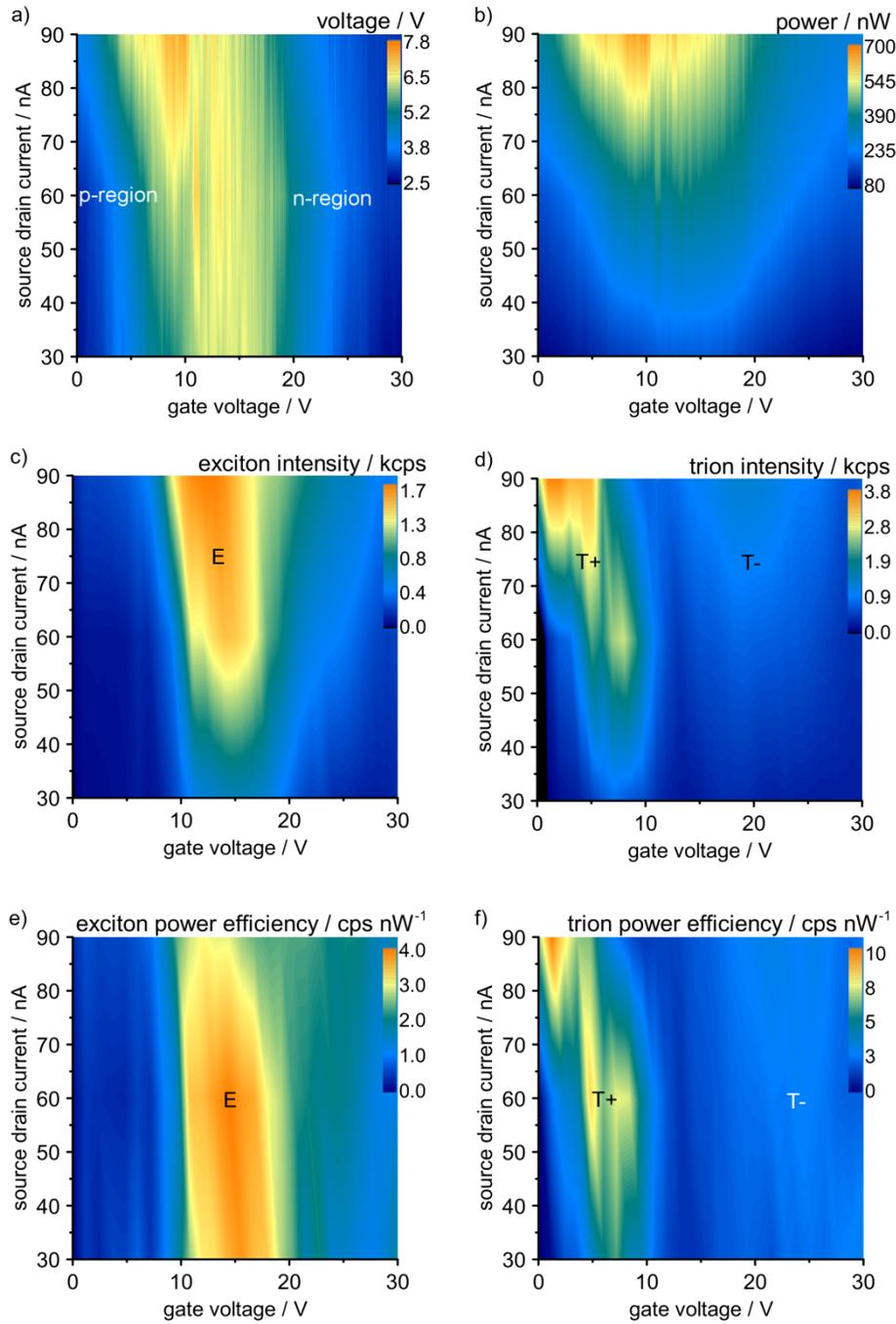

Figure S2: Electrical biasing and power dissipation versus light emission from excitons and trions in a hole-doped (9,8)-device. (a-d) Voltage, electrical power, and integrated exciton and trion intensity versus current bias and gate voltage (exiton range: 1370-1500 nm; trion range: 1500-1613 nm). (e-f) Exciton and trion power efficiency (photon flux / electrical power). The positively charged Trion (T$^+$) is more intense than the negatively charged Trion (T-) and appear at negative and positive gate voltages with respect to the excitonic emission (E).

**Impact of annealing on transconductance curve**

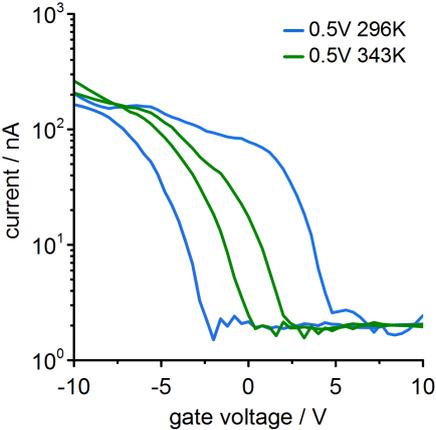

Figure S3: Transconductance curves measured in vacuum. The large hysteresis measured at 296 K reduces upon annealing at 343 K.

**Carbon nanotube electroluminescence detection efficiency**

To determine the efficiency of converting charges into photons for a carbon nanotube device, we need to know the sensitivity of the setup to photons emitted by the nanotube. We approach the problem by determining $f_{tube \rightarrow objective}$, the fraction of emitted photons collected by the microscopy objective, and $f_{countrate \rightarrow flux}$, the factor that converts detector count rate into photon flux. The total setup efficiency $\eta_{setup}$ is then given by

$$\eta_{setup} = f_{tube \rightarrow objective} \cdot f_{countrate \rightarrow flux} \qquad \qquad (eq.\ 1),$$

in units of detector count rate per emitted photon flux.

<u>(A) Fraction of emitted photons collected by the microscopy objective ($f_{tube \rightarrow objective}$)</u>

$f_{tube \rightarrow objective}$ is determined by the numerical aperture (NA) of the objective and the radiation pattern of the nanotube emitter. The latter depends in the position and orientation of the nanotube with respect to the layered substrate and its orientation to the objective. In the setup the sample is mounted in such a way that the surface normal of the SiO₂/Si substrate is parallel to the optical axis of the microscope objective. We define this direction as the z-axis and use the ISO spherical coordinate system with the radial distance r, polar angle θ, and azimuthal angle φ.

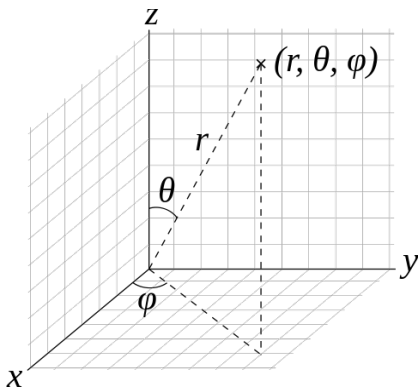

Figure S4a: Spherical coordinate system. The optical objective is parallel to the z-axis. We consider the nanotube aligned within the x-y-plane along the x-axis (p∥x) or the y-axis (p∥y).

In a first step we model the carbon nanotube as an electrical dipole in free space, a problem that can be easily solved analytically in 3D. The radiation power pattern $f_P$ follows[1]

$f_P = \cos^2(\theta) \cdot \cos^2(\varphi) + \sin^2(\varphi)$,  for p ∥ x  (eq. 2.1);

$f_P = \cos^2(\theta) \cdot \sin^2(\varphi) + \cos^2(\varphi)$,  for p ∥ y  (eq. 2.2);

$f_P = \sin^2(\theta)$,  for p ∥ z  (eq. 2.3).

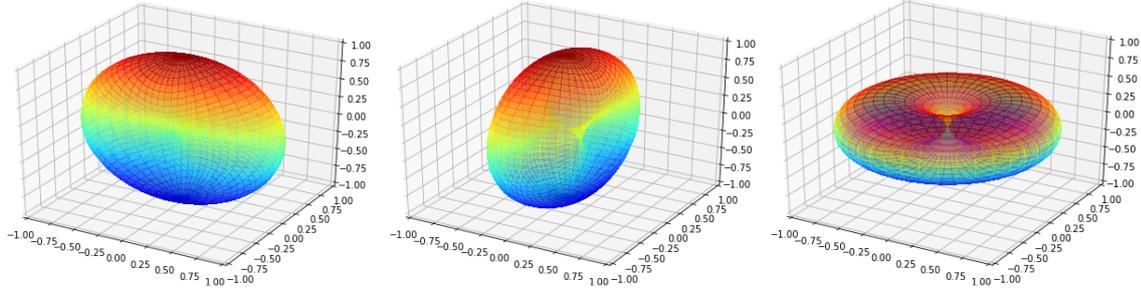

Figure S4b: Radiation pattern for a free-space dipole with orientations p ∥ x (left), p ∥ y (middle), p ∥ z (right). The coordinate system is identical with Fig. S4a.

In our experiment the nanotube is oriented within the x-y-plane, corresponding to the equivalent cases p∥x and p∥y. The collection efficiency $f_{\text{tube}\rightarrow\text{objective}}^{free-space}$ is then calculated by integrating $f_P$ over the acceptance angle of the objective, normalized by the integral over all angles:

$$f_{\text{tube}\rightarrow\text{objective}}^{free-space} = \int_{\theta=0}^{\text{asin}(NA)} \int_{\varphi=0}^{2\pi} f_P(\theta,\varphi)\sin(\theta)d\theta d\varphi \; / \; \int_{\theta=0}^{\pi} \int_{\varphi=0}^{2\pi} f_P(\theta,\varphi)\sin(\theta)d\theta d\varphi \qquad (eq. 3)$$

For NA = 0.6 we obtain 13.6 % for p∥x and p∥y, and 2.8 % for p∥z. For comparison, an isotropic emitter gives 10 %.

The influence of the SiO$_2$/Si substrate on $f_{\text{tube}\rightarrow\text{objective}}$ we first targeted numerically. We have used a finite difference time domain solver (Lumerical) to calculate the radiation pattern in a 2D simulation space (z-x-plane). The simulation space is confined by absorbing phase matching boundary layers (PML) and the lateral space was chosen large enough such that near fields that are propagating close to the surface are captured by the near-field linear transmission monitor before getting absorbed at the PML boundaries (x-span: 16000nm, z-span: 2000nm). Refractive index values for the 300 nm thick SiO$_2$ layer and the Si were taken from the Lumerical standard database (Palik). To save computation time we refrained from placing the dipole exactly at the air/SiO$_2$ interface (the surface) and performed simulations with the electrical dipole 20 nm below the surface and 3nm above the surface. The linear near-field transmission monitor was placed 10nm above the surface. The far-field radiation power distribution was obtained by projecting the transmitted near-field power on to a 1m radius semi-circle. Figure S4c shows the radiation pattern in the wavelength range 1000 - 1600 nm for the dipole in p‖y orientation. The values for the dipole above and below the surface were normalized with the corresponding homogeneous medium simulations.

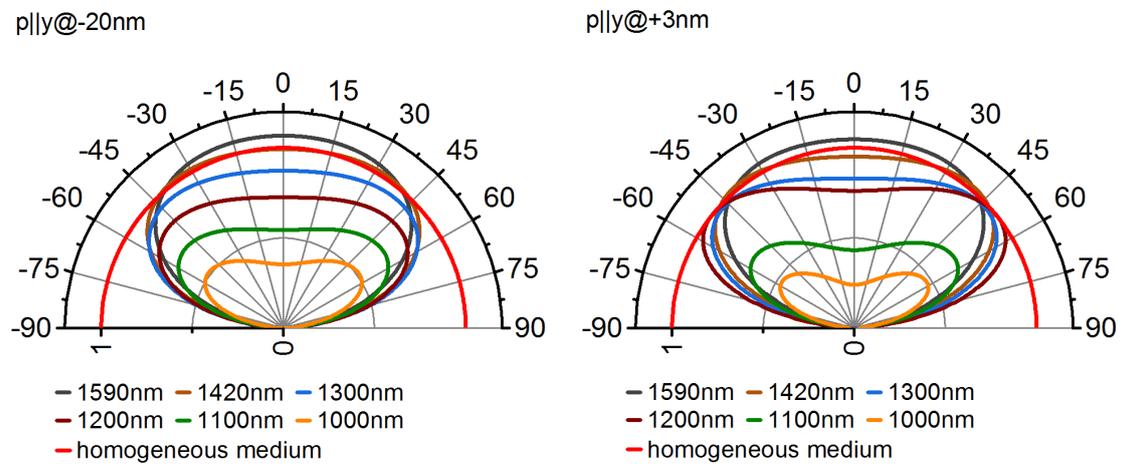

Figure S4c: Radiation pattern as a function of emitter wavelength of an electrical dipole in p‖y orientation, located 20 nm below (left) and 3nm above (right) the SiO$_2$/Si substrate surface.

The results are similar for the emitter above and below the surface for wavelengths above 1300nm and shows that under conditions where waveguide modes and surface plasma polaritons are absent the distance to the surface is not critical.

Next, the fraction of radiation that is collected by the NA=0.6 objective was obtained by integrating the radiated power over the acceptance angle of the objective, normalized to the total emitted power. The results are shown in Figure S4d. It is important to note that the 2D simulations yield different results for the p‖x and p‖y orientations, whereas these orientations are equivalent in 3D. For dipoles in free space we have reproduced the 2D simulations analytically (p‖x:35.8%, p‖y:20.5%) and compared them to analytical 3D result (p‖x=p‖x=13.6%) (dashed lines in Fig. S4d). The results show that 2D simulations are not suitable to map out properly the 3D problem. We have therefore written a python code for the analytic 3D solution for a dipole on a layered structure following chapter 10 in the Principles of Nano-Optics book of Novotny and Hecht.[2] The code was verified with the kind assistance of Lukas Novotny. The full green line in Fig. S4d gives now the correct result for the fraction of light that is collected by a NA=0.6 objective for a dipole emitter in p‖x or p‖y orientation on 300nm SiO$_2$/Si. At 1420 nm we collect 18.79% of the emitted light, whereas at 1000 nm it is only 5.83%. Compared to the 13.6% in a homogeneous medium, we collect in the near-infrared more light due to constructive interference from the layered substrate.

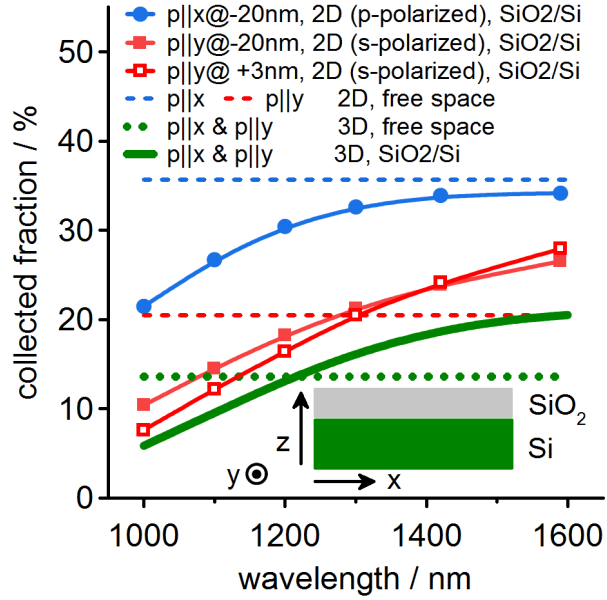

Figure S4d: Fraction of emitted photons collected by the NA=0.6 microscopy objective as a function of wavelength for the two dipole orientations (p∥x, p∥y), 3nm above and 20 nm below the SiO$_2$/Si substrate surface. The 2D FDTD simulations are compared to 2D free space simulations and to the analytical 3D result using equation 3. The full green line is the correct, analytical 3D solution of the problem.[2]

Hence, at NA = 0.6 and λ = 1400 - 1600nm, $f_{tube->objective}$ ≈ 19 - 20% for a nanotube on a 300 nm-SiO$_2$/Si surface.

### (B) Factor converting detector count rate into photon flux ($f_{countrate->flux}$)

We have experimentally determined $f_{countrate->flux}$ with a calibrated Ocean Optics HL-3P-INT-CAL-EXT Vis-NIR light source for integrating spheres. The calibration data is provided by the manufacturer as radiant flux in the physical unit μW/nm with an uncertainty below 10%. This light source was connected to an SM1 input port of a Thorlabs IS200 integrating sphere (2″ inner diameter). The 1mm output port of the sphere was covered with a 5 μm diameter pinhole, to mimic a point like Lambert emitter. The flux $\phi_{out}$ emitted through the pinhole corresponds to the radiant flux of the calibrated source $\phi_{in}$ attenuated by β = $\phi_{in}/\phi_{out}$. The factor β is determined by[3]

$$\beta = \frac{A_{pinhole}}{\pi \cdot A_{sphere}} \cdot \frac{\rho}{1-\rho(1-f)} = \frac{A_{pinhole}}{A_{sphere}} \cdot M \qquad \text{(eq. 4).}$$

M is the sphere multiplier factor which depends on the sphere reflectance ρ and the port fraction f = ($A_{input}$ + $A_{output}$)/ $A_{sphere}$, the ratio of the sum of the input and output ports and the sphere area. M was determined experimentally to be 31, and with $A_{pinhole}$ = π·(5/2·10$^{-6}$m)$^2$ and $A_{sphere}$ = 4π ·(25.4·10$^{-3}$m)$^2$ we obtain β = 7.5·10$^{-8}$.

The pinhole was brought into the focus of a Zeiss LD-Plan Neofluar (40x/NA=0.6) objective. A Thorlabs FELH1000 longpass filter was inserted to block visible light that could reach the NIR detector by second order diffraction. The light was dispersed with a 85 G/mm, 1.35 µm blazed grating in an Acton SP-2360 (f/3.9) imaging spectrograph (center wavelength set to 1350 nm), and spectra were recorded with a Princeton Instruments Pylon-IR 1.7 detector. The detector was operated at -100 °C and at high-gain/500kHz setting. Under these conditions the factor to convert photon flux into detector count rate is given by

$$f(\lambda)^{pinhole}_{countrate \to flux} = \frac{\beta \cdot calibration\ spectrum\ [\mu W/nm] \cdot 10^{-6} \cdot \lambda/hc}{measured\ spectra\ [count\ per\ second]}$$  (eq. 5).

The factor $10^{-6}$ accounts for the calibration spectrum given in µW and $\lambda/hc$ converts power into photon flux.

Figure S4e shows the wavelength dependent conversion factor $f(\lambda)^{pinhole}_{countrate \to flux}$ together with the calibration spectrum and the spectra measured with the pinhole.

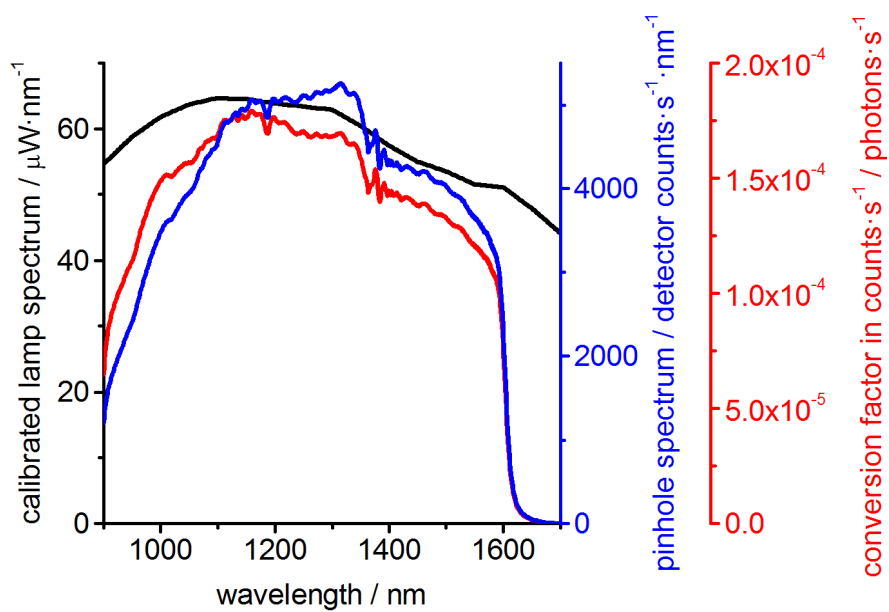

Figure S4e: Conversion factor $f(\lambda)^{pinhole}_{countrate \to flux}$ (red) calculated via eq. 5 from the calibrated lamp spectrum (black) and the measured pinhole spectrum.

Dividing the red curve by $f(1430nm)^{pinhole}_{countrate \rightarrow flux}$ = 1.4·10⁻⁴ counts·s⁻¹/ photons·s⁻¹ gives the correction curve for the relative spectral sensitivity

$$f(\lambda)_{relspectcorr} = \frac{f(\lambda)^{pinhole}_{countrate \rightarrow flux}}{f(1430nm)^{pinhole}_{countrate \rightarrow flux}} \qquad \text{(eq. 6)},$$

which has been used to normalize every electroluminescence spectrum in this work.

We can verify $f(1430nm)^{pinhole}_{countrate \rightarrow flux}$ by adding up the known losses, gain factors and efficiencies of the individual components:

$$\eta_{setup} = QE_{InGaAs}/f_{gain} * \eta_{grating} * R_{mirror} * T_{objective} * f_{pinholeemitter \rightarrow objective} * f_{losses} \text{(eq. 7)}$$

With the quantum efficiency of the detector $QE_{InGaAs}$=0.5; gain factor of the detector $f_{gain}$=1/75; grating efficiency $\eta_{grating}$=0.6·0.96·0.99; reflective mirror losses $R_{mirror}$ = 0.97; transmission of the objective $T_{objective}$=0.35; and the fraction of light from the Lambertian emitter collected by the objective $f_{pinholeemitter \rightarrow objective}$ =0.288 (see below). To match the experimental result we need to postulate additional beam losses $f_{losses}$~60%.

The value $f_{pinholeemitter \rightarrow objective}$ =0.288 was derived following the work of Tang et al.[4] For a 5 µm diameter pinhole at the center-top position of a 3mm diameter wide and 6mm long exit port, $f_{pinholeemitter \rightarrow objective}$ = 0.8·(NA)². The wanted factor $f_{countrate \rightarrow flux}$ in eq. 1 is then given by

$$f(\lambda)_{countrate \rightarrow flux} = \frac{1}{0.29} \cdot f(\lambda)^{pinhole}_{countrate \rightarrow flux} \qquad \text{(eq. 8)}$$

and finally for λ > 1300nm we can determine the total sensitivity of the setup to photons emitted by the nanotube as

$\eta_{setup}$ = $f_{tube \rightarrow objective}$ · $f_{countrate \rightarrow flux}$ = 0.19·(0.29)⁻¹·1.4·10⁻⁴ = 9.2·10⁻⁵ counts·s⁻¹/ photons·s⁻¹ (eq. 1)

For low-temperature experiments the losses at the cryostat window have to be taken in to account (95% transmission). **The final result is then $\eta_{setup}$ = 8.7·10⁻⁵ counts·s⁻¹/ photons·s⁻¹, which means each detector count corresponds to ≈ 11500 photons emitted by a nanotube located on a 300nm-SiO₂/Si substrate.**

Since the spectra shown in the main text have already been normalized with the relative spectral sensitivity curve $f(\lambda)_{relspectcorr}$, all it takes to convert detector counts per second into photon flux is multiplying the spectra for all wavelengths with the factor $\eta_{setup}$ = 8.7·10⁻⁵ counts·s⁻¹/ photons·s⁻¹.

**Electroluminescence quantum efficiency ELQE**

The electroluminescence efficiency ELQE can be defined as relation between the number of emitted photons and charges, passing through the nanotube which yields for excitons (for trions 3e):

$$\eta_{ELQE} = \frac{N_{photon}}{N_{charges}} = \frac{I(cps)}{I(A)} \frac{2e}{\eta_{setup}} \qquad \text{(eq. 9)}$$

with $\eta_{setup}$ determined above, which gives a conversion factor of

$$\eta_{ELQE} \approx 3 \cdot 10^{-6} \frac{I(cps)}{I(nA)} \qquad \text{(eq. 10)}$$

The power efficiency is the relation between emission and electrical power:

$$\eta_{Power} = \frac{P_{photon}}{P_{charges}} = \frac{I(cps)}{I(A)} \frac{E_{(9,8)}(eV)}{V} \frac{2e}{\eta_{setup}} = \eta_{ELQE} \frac{E_{(9,8)}(eV)}{V} \qquad \text{(eq. 11)}$$